\documentclass{article}

\usepackage[english]{babel}

\usepackage[letterpaper,top=2cm,bottom=2cm,left=3cm,right=3cm,marginparwidth=1.75cm]{geometry}

\usepackage{amsmath}
\usepackage{graphicx}
\usepackage{xargs}
\usepackage{enumerate}
\usepackage[colorlinks=true, allcolors=blue]{hyperref}
\usepackage{xcolor}  

\usepackage{hyperref}
\usepackage[
    backend=biber,
    style=authoryear,
    dashed=false
]{biblatex}

\newcommand{\hmpc}{\,$h^{-1}$\,Mpc}
\newcommand{\hgpc}{\,$h^{-1}$\,Gpc}

\newcommand\citeposs[1]{\citeauthor{#1}'s\ (\citeyear{#1})}
\newcommand{\footremember}[2]{%
    \footnote{#2}
    \newcounter{#1}
    \setcounter{#1}{\value{footnote}}%
}

\title{Contemporary Philosophical Perspectives on the Cosmological Constant}
\author{Adam Koberinski\footremember{Pitt}{Center for Philosophy of Science, University of Pittsburgh, Pittsburgh, PA, USA} \footnote{adam.koberinski@pitt.edu}\, , Bridget Falck\footnote{bridget.falck@gmail.com}\, , and Chris Smeenk\footremember{Western}{Department of Philosophy and Rotman Institute of Philosophy, University of Western Ontario, London, ON, Canada} \footnote{csmeenk2@uwo.ca}}

\begin{document}
\maketitle

\begin{abstract}
The (re)introduction of $\Lambda$ into cosmology has spurred debates that touch on central questions in philosophy of science, as well as the foundations of general relativity and particle physics. We provide a systematic assessment of the often implicit philosophical assumptions guiding the methodology of precision cosmology in relation to dark energy. We start by briefly introducing a recent account of scientific progress in terms of risky and constrained lines of inquiry.  This allows us to contrast aspects of $\Lambda$ that make it relevantly different from other theoretical entities in science, such as its remoteness from direct observation or manipulability. We lay out a classification for possible ways to explain apparent accelerated expansion but conclude that these conceptually clear distinctions may blur heavily in practice. Finally, we consider the important role played in cosmology by critical tests of background assumptions, approximation techniques, and core principles, arguing that the weak anthropic principle fits into this category. We argue that some core typicality assumptions---like the Copernican principle and the cosmological principle---are necessary though not provable, while others---like the strong anthropic principle and appeals to naturalness or probability in the multiverse---are not similarly justifiable. 
\end{abstract}


\section{Why \textit{Philosophy} of $\mathrm{\Lambda}$?}

In the last two decades, cosmologists have made an increasingly strong observational case in favor of a substantial contribution to the total mass-energy budget from an effective cosmological constant term ($\Lambda$), also called ``dark energy.''  Despite reaching consensus that this term contributes about $70\%$ of total mass-energy, many cosmologists regard this discovery as deeply mysterious:  we do not understand the underlying physics.  And the most widely accepted proposal for the underlying physics, that $\Lambda$ arises from vacuum energy of quantum fields, turns the mystery into a crisis:  (admittedly na\"ive) calculations lead to a value $120$ orders of magnitude too large!  Furthermore, corrections to the na\"ive calculations would be expected to force $\Lambda$ to be exactly zero---closer to the observed value, but now erring on the other side by being too low.  These problems---called the old and new cosmological constant problems, respectively---have generated a vast literature.  There are many responses to the problems, embodying quite diverse theoretical approaches, yet no clear consensus.    

Our goal in this paper is to provide scientists with contemporary philosophical perspectives which can be used to evaluate, assess, and contextualize issues related to the cosmological constant and dark energy. We make no claim to providing innovative contributions to the analysis of observations or cosmological theory. Rather, we will carefully scrutinize what some consider to be a crisis in cosmology in potentially unfamiliar---but hopefully helpful!---terms based on contemporary philosophical thought in the epistemology and methodology of physics. We aim to introduce a wider array of philosophical tools than are available in a  na\"ive and static view of how science works.

We will first address methodological issues that arise in relation to the cosmological constant. What are the steps involved in the production of scientific knowledge? What are the allowed moves as we collect data and build theoretical models, and are some moves preferred over others? How do we judge scientific progress? Each scientific discipline answers these questions in different ways according to what best suits its subject matter and suite of techniques, and we will focus on the aspects that are germane to cosmology. However, there are a few general principles important to understanding good science that are adapted and refined in each case. First, science is ultimately empirically grounded, despite the inability to test observations in a completely theory-neutral way. Even though the theoretical and empirical are not so easily disentangled---we can't separate data from our ideas about data entirely, and there are models at every stage of the measurement process---empirical evidence is the final arbiter of epistemic warrant \parencite{McIntyre,sep-theory-observation}. 
Next, a scientific discipline is an iterative enterprise whose aim is to build a more detailed, complete picture of its domain of phenomena over time. The successive buildup is predicated on the theoretical background that has come before: good science is risky in proposing new observable consequences, and it is constrained to minimally modify the existing well-confirmed framework. Science is both a rational and social enterprise as well: a scientific community has shared rules and ideals, and it is the healthy disagreement of experts that leads to consensus, over time. Indeed, given the inevitable influence of the individual and social values of the practitioners of science, it is the ``subjection of hypotheses and theories to multivocal criticism that makes objectivity possible'' \parencite[p. 213]{Longino}.  In Section~\ref{Methodology}, we will frame debates about the cosmological constant as disagreements over what methodological steps are appropriate and will best lead toward a fruitful line of inquiry.

We next address scientific explanation in Section~\ref{Explanation}: the art and challenge of connecting data collected from many varied experiments and observations into a coherent whole, learning from those data the regularities or physical laws that guide how the universe works, and representing that knowledge in a set of theoretical models---not necessarily in that order. Which sorts of things demand an explanation, and which can we take instead as contingent facts? What kinds of explanations are possible, and what kinds of explanation should be sought? When faced with competing explanations, how do we judge which one is more suitable or more successful as an explanation, if at all we can? Scientific explanation is a rich topic in philosophy of science \parencite{SEPExplanation}, and we make no pretense to an exhaustive overview of ideas. Instead, we present our own view on explanation as it pertains to understanding accelerated expansion and dark energy.

There are many explanations offered for what $\Lambda$ could be in the literature. When the observational evidence for accelerated expansion was new, there were more explanations offered at the level of the data: the data were evidence of calibration or other measurement errors, the luminosity evolution of supernovae, the presence of grey dust, or other astrophysical effects. After gathering evidence from multiple independent probes of the expansion rate, the community has mostly settled on the consensus that the expansion of the universe is accelerating, and explanations are sought that apply across diverse sources of observational evidence. 
Broadly, we can put explanations of $\Lambda$ into 6 categories that describe the cosmological constant term as:
\begin{enumerate}
    \item ... a matter-sourced cosmological constant, explained by some modification to the Standard Model of particle physics.
    \item ... a true constant of nature---a free parameter we have measured the value of, like the gravitational constant $G$.
    \item ... the effect of dark energy, a new type of mass-energy that may be a scalar field like the Higgs particle, or something more complicated, and is not necessarily a constant.
    \item ... the effect of a correction to the equations of general relativity brought on by some more complete theory of gravity, with the result that it looks like a scalar field but has dynamical implications as well.
    \item ... an effect of the misapplication of the FLRW metric in what is demonstrably a bumpy, locally inhomogeneous universe, or an effect some other systematic errors in our modelling of phenomena related to accelerated expansion.
    \item ... the vacuum energy in our section of the multiverse, which can have different values than the one we expect from particle physics due to some unknown mechanism, and its value is explained anthropically.
\end{enumerate}
All of these possible explanations are allowed by the ``rules of the game'', but depending on ones' perspective, some of these might not even be considered explanations at all---particularly the first two. The first is certainly not an explanation, yet; it is the promise of an explanation to come. But since science is a process that takes time, it may turn out to be the solution. The second may not be an explanation at all but rather the position that no explanation is needed. To some, the difference may be moot. Since phenomena are understood in the context of a theoretical framework, the same phenomenon explained in one framework may not need to be explained with respect to a different framework, and what counts as an explanation in one may not be a satisfactory explanation in another. In Section~\ref{Explanation}, we lay out different strategies for explaining new phenomena in physics, placing common approaches to explaining the cosmological constant in the context of these strategies. We then discuss what we call eliminative reasoning, the phenomenological approach that attempts to build an observational case that accelerated expansion cannot be explained by a true cosmological constant. Finally, we turn to the thorny problem of being unable to distinguish between potential explanations, which is known as underdetermination \parencite{sep-scientific-underdetermination} and which may be a particular problem for explanations involving dark energy (number 3 above) and modified gravity (number 4).

Underdetermination is not always a threat to scientific progress, but in some sense, it never goes away. There may always be some source of error, an unknown systematic effect, a bug in a data pipeline, a loose cable, an undiscovered force of nature, you name it. The day-to-day practice of science is in coming up with potential sources of error, developing ways to find them or rule them out as insignificant, and implementing those tests \parencite{Mayo1996}. The theoretical framework of a scientific discipline has been tested and vetted enough to be trusted, but new data or new ideas may discover parts of the framework that are unstable. Thus some explanations, like number 5 above, are attempts to refine or readjust the approximations and modeling assumptions of the theoretical framework. In Section~\ref{Assumptions} we examine the foundational assumptions in the standard model of cosmology that relate to typicality, namely the Copernican and cosmological principles, and the related issue of selection effects given our limited vantage point from which to study the Universe. We take a critical view of strong anthropic explanations (number 6 above), but we argue that weak anthropics help point out similar selection bias effects. Finally, we turn to naturalness arguments and the ``prediction'' of $\Lambda$ using anthropic reasoning, arguing that the prediction should be instead thought of as a plausible constraint motivating a solution to the cosmological constant problem.

We think that contemporary philosophy of science has developed sophisticated tools for understanding these issues in science generally, and our goal is to introduce these tools to cosmologists thinking about dark energy and the status of $\Lambda$ within the $\Lambda$CDM model.

\section{The methodology of cosmology}
\label{Methodology}

Cosmologists typically take $\Lambda$CDM as the basis for further research, akin to the first step in a series of approximations that hopefully converge towards ever greater detail and fidelity.  Yet several of its features raise challenges to this view.  Recently high-precision measurements of the parameters of the $\Lambda$CDM model have revealed systematic discrepancies, with that between ``local'' and ``global'' measurements of $H_0$ sparking considerable interest and debate.\footnote{See, for example, \textcite{di2021realm} for a recent review of proposed solutions, and \textcite{smeenktrouble,gueguen2022crack} for philosophical discussions.}  This line of inquiry may force profound revisions of $\Lambda$CDM if these discrepancies persist as sources of systematic uncertainty are brought under control, a possibility we will set aside for this paper. The need for a non-zero $\Lambda$, the focus of our discussion, is also often taken to count against $\Lambda$CDM.\footnote{A true cosmological constant does not vary throughout spacetime, so $\Lambda$ would in this sense be a universal parameter. We follow common practice in using ``dark energy'' to describe a dynamical contribution that can be well-approximated by a non-zero $\Lambda$ in Einstein's field equations, but can vary slowly at large spatio-temporal scales.} Should we regard $\Lambda$, a free parameter to be fixed by observations, as an \textit{ad hoc} modification to save the standard model, and methodologically unsound? Does the need to introduce new, otherwise undetected types of matter and energy ($\Lambda$ as well as dark matter) reflect striking discoveries, or is it instead the sign of a degenerating research program?  Philosophers have long aimed to provide guidance on such questions, based on elucidating good methodology and showing how it contributes to progress.  These questions are challenging due to the distinctive aspects of cosmology compared to other domains of physics.\footnote{George Ellis has returned to this theme periodically, see for example \textcite{ELLIS2014}, as well as \textcite{goenner2010kind}.  Recently some cosmologists have turned to the ideas of mid-twentieth century philosophers such as Popper, Kuhn, and Lakatos to answer questions regarding the appropriate method \parencite{merritt2020philosophical}.}  While we applaud this philosophical turn, on our view more recent work in philosophy of science supports a richer and more insightful analysis.  Drawing on what we take to be the best current accounts of scientific methodology, we will contextualize debates regarding $\Lambda$ as disagreements over the best methodology for dealing with anomalies in cosmology, the most preferred types of explanations, and ways to deal with lack of direct evidence. These are debates over the ``rules of the game'' of inquiry into the large-scale features of our universe.

\citeposs{Popper} original account of scientific methodology was a two stage process: start with a hypothesis, and then seek to refute it. Observation was the key pillar, an objective arbiter of refuted or corroborated hypotheses. Views like Popper's are examples of empiricist methodology of science. Since \citeposs{Kuhn} landmark publication of \emph{The Structure of Scientific Revolutions}, however, philosophers have had to confront and reconcile traditional empiricist views of science with the reality that observation and hypothesis testing are inherently theory-laden. Our best theories provide the concepts, methods, and rules for investigating the natural world. We fit nature into the conceptual scheme provided by a theoretical framework, and progress is made the more we can subsume. Rather than being passive observers that directly experience reality, we must grapple with the idea that humans play an active role in perception. These facts have proven hard to reconcile with naive accounts of empiricism like Popper's, where hypotheses can be tested in isolation, and theories can be tested, falsified, jettisoned, and replaced. However, to give up on objectivity and the priority of empirical results seems to us too strong a move in the direction of Kuhn's metaphysical constructivism. Empirical grounding is still a major hallmark of successful science.

Theories often transform our understanding of a domain of phenomena, in a way that bypasses or resolves old problems but also generates new ones.  An account of progress should help to distinguish between transformations that are steps in the right direction and those that are detours into dead ends.  Furthermore, any account of scientific progress has to show how theoretical transformations are compatible with, or even essential to, the accumulation of empirical knowledge, as well as clarifying the nature of the knowledge we have achieved.  Though the basic empiricist idea that experience is the ultimate authority has merit, we need a more nuanced account of progress in science than that offered by Popper and his contemporaries to answer such questions. 

On our view, the proper unit of analysis for an account of scientific progress is a line of inquiry extending over time, an activity guided by theory, rather than a theory considered at an isolated moment. We should not frame the question of theory evaluation as merely that of determining whether a theory is compatible with a body of available data.  This threatens to downplay just how much work is needed, both in proposing and developing new experiments and understanding what the theory implies, to enable such a comparison. In addition, this work is thoroughly theory-mediated:  observations or experiments are designed to target quantities introduced by the theory, their reliability is assessed based on the theory itself, and so on. We need to presuppose some aspects of the theory itself to bring it into contact with nature. The second problem with this framing is that it misconstrues the nature of success.  Scientists do not only seek theories that are compatible with a given body of data, at a moment;  rather, they seek a theory that can guide inquiry fruitfully as the available data steadily increases in scope and precision. There is a compelling and quite intuitive contrast between theories that succeed as increasingly high quality data comes in, by (apparently) getting the relevant dynamics or other important dependencies right, as opposed to theories that succeed in a much less satisfying sense---they ``merely'' fit the data, by exploiting theoretical flexibility.  Mere curve fits do not support ongoing inquiry.       

Inspired by \textcite{stein1994some,smith2010revisiting,smith2014closing} (and at the risk of drastically oversimplifying), we propose that progressive lines of inquiry have two essential features:  they are both risky and constrained.  A line of inquiry is risky in so far as new data, of increasing scope and precision, could prove to be in conflict with the underlying theory; it is constrained in so far as there are only a few available responses, determined in advance, to any such conflicts.  Any line of inquiry in physics includes the construction of increasingly detailed descriptions of some target phenomena, and it is almost always the case that at any given stage of development there are discrepancies between theory and observation driving further refinements. 
Experiments and observations of the target system are rarely designed to test the theoretical framework by directly ``checking predictions.''  Instead, the focus is often on characterizing the discrepancies remaining at a particular stage of modeling with sufficient precision to help identify possible physical sources. The framework makes it possible to evaluate how adding these sources would contribute to observed behaviour, and hence to see whether they would resolve the original discrepancy. In this way, discrepancies are often better sources of evidence for a line of inquiry than finding simple agreement. Iterating this process generates a series of increasingly detailed approximations, with new details added at each stage and with further---hopefully smaller---discrepancies identified at later stages. The basic theoretical assumptions are constantly put to the test as the more sophisticated models are developed and compared to nature.  The theoretical framework guides the design of new tests, providing the structure needed to organize and interpret further findings. 
 
Given the thorough theory-dependence of this account, it is natural to worry that any given line of inquiry might be shielded from rejection.  Yet relying on a framework to guide and make sense of empirical results can still be risky.  Acknowledging that this process is heavily theory-mediated does not imply that ``anything goes.''  For example, theories often make it possible to bring multiple lines of evidence to bear on a specific feature, such as the value of a particular parameter. Yet nothing guarantees that these different measurements will agree, if these techniques are truly independent of one another. Hence consistency among diverse measurements provides evidence in favour of the relevant parts of the theory. The degree of risk-taking depends on how tightly constrained the line of inquiry is. If the theoretical framework can be modified freely, on a case-by-case basis, to account for isolated discrepancies, then there really is no risk associated with accepting it.  Successful anomaly resolution then provides evidence for little more than the ingenuity of theorists.  The crucial point is that theory-dependence itself is not an obstacle to developing strong evidence; challenges instead stem from vagueness, flexibility in applications, and lack of clear constraints. 

Celestial mechanics provides a concrete illustration of this rather abstract characterization of successful theory-mediated research. \textcite{smith2014closing} describes the centuries long line of inquiry based on treating solar system motions in terms of Newtonian gravity in these terms.  This line of inquiry proceeded by identifying discrepancies between current models and observations, searching for new gravitational sources, refining approximation methods to better model the complex multi-body gravitational forces acting on each planet in the solar system, and so on. This allowed for an increase in precision of both prediction and observation, which in turn allowed for the discovery of new physically relevant causal dependencies in the motion of the planets.  The assumption that all of these details could be accounted for in terms of the gravitational interaction was risky.  In almost all cases, this risky assumption paid off, leading to direct detection of physical sources for these aspects of celestial motions---for example, new planets, or the varying rotational speed of the Earth. The line of inquiry was also constrained:  certain types of changes---like changing the force law, or positing new, non-gravitational forces---were almost completely excluded from the program of Newtonian celestial mechanics, and were only considered as a last resort when all other means were exhausted.\footnote{Non-gravitational forces did come into play in a few important cases, such as the explanation of the secular acceleration of the moon as a consequence of the Earth's rotation slowing due to tidal friction.} Eventually this line of inquiry led to minute discrepancies that could not be addressed by changes compatible with the constraints, and these true anomalies forced the revision of core principles of Newtonian mechanics. 

In analyzing contemporary science, we lack the benefit of hindsight in deciding what qualifies as a progressive line of inquiry.  But we can still ask whether a particular line of inquiry---specifically, that based on the standard model of cosmology---is appropriately risky and constrained. What are the rules of the game cosmologists have adopted, perhaps implicitly, in developing ever more detailed models? For the reintroduction of $\Lambda$, do the rules give unambiguous answers?

Often debates regarding the introduction of $\Lambda$ focus on one dimension of this assessment:  what are the appropriate constraints on cosmology? On the one hand, the addition of dark energy (and dark matter) to the big bang model can be seen as a natural evolution of the framework.  Dark energy in the form of a cosmological constant already appears in the underlying dynamics, and its value can be constrained by several different types of observations---inferred from the CMB, structure formation models, and local observations of accelerating expansion. On this view, one can think of cosmology as allowing us to discover new exotic types of matter and energy not yet detected via non-gravitational means. It would be rash to restrict cosmological models to familiar types of matter.  Just as astronomers used gravitational theory to reveal new planets, astrophysicists have discovered that a variety of new states and types of matter are needed to model extreme environments.  On the other hand, detractors can point to the need to introduce supposedly \textit{ad hoc} new ingredients as a sign that the cosmology framework is woefully under-constrained.  Without dark energy and dark matter, the same observations taken to vindicate $\Lambda$CDM would decisively refute it. On this view, a more plausible and constrained version of cosmology does not permit introducing matter-energy sources with properties so different from that of ``ordinary'' matter because this would only serve to hide the flaws of the standard model.

On our view, a more convincing assessment has to analyze the interplay between risk and constraints over the long term. In the case of contemporary cosmology, just as with celestial mechanics, the era of precision cosmology has been made possible by adopting $\Lambda$CDM. $\Lambda$CDM provides a theoretical framework upon which we predicate searches for new physical effects relevant to the dynamics and structure of the large-scale degrees of freedom of the universe. Implicit in this framework is a methodology for interpreting our observations in terms of FLRW-like spacetimes with large scale (near) homogeneity and isotropy. These are some of the clear constraints imposed as rules for cosmology. Based on this framework, cosmologists have identified the dominant causal factors in the dynamical evolution of the universe, arguably including the discovery of dark matter and dark energy, along with more subtle effects (e.g., gravitational waves). Most cosmologists accept that the $\Lambda$CDM model serves, as with the Newtonian account of celestial mechanics, as at least a good first approximation.  If that is correct, then even if the $\Lambda$CDM model is eventually replaced, the physical details that it has enabled us to discover should be preserved in any successor theory. 
 
Assessing the degree of risk associated with accepting $\Lambda$ is particularly challenging for two different reasons.  The first is ambiguity in what it means, precisely, to accept $\Lambda$:  different interpretations of what $\Lambda$ is lead to diverging paths for further work. This is where ambiguity over the rules of the game comes in. How risky is $\Lambda$, i.e., how many new phenomena should we expect from positing something approximating a cosmological constant? Relatedly, how constrained is the form for new physics underlying $\Lambda$? We might treat $\Lambda$ as simply a free parameter appearing in Einstein's field equations \parencite{bianchi2010all}. Here there is little else we would expect to observe, but the form of physics is tightly constrained to be a true cosmological constant. The question is then whether we can consistently determine the value of $\Lambda$; since the value itself is merely contingent, there would be no further lines of inquiry.  Analogously, we no longer seek an explanation for the number of planets in the solar system, which Newtonian astronomy treats as a contingent, unexplainable fact.  But, by contrast, we might instead regard the apparent $\Lambda$ term as resulting from the dynamical features of an entirely new kind of matter, dark energy. This line of reasoning would be more risky, but less constrained. Though there is significant freedom in tuning the new physics, in principle one could pursue a line of inquiry devoted to learning more about the underlying physics.  We will return to these issues in further detail in Section~\ref{Explanation} below.
 
The second challenge stems from the inaccessibility of the scales where the effects of $\Lambda$---and the risks associated with accepting a particular way of understanding it---become manifest. Thus it is unclear how risky certain candidate explanations actually are in practice. In many other domains of science, introducing a new entity leads to a variety of risky predictions. For example, accepting the Higgs boson in the Standard Model led to a number of consequences that were eventually confirmed by the LHC: first, a particle consistent with a spinless boson was discovered with a mass of about 125GeV; then, with further production experiments its properties were more accurately determined to match the `vanilla Higgs'. Even when manipulability or direct experimentation is lacking, one can still look for independent evidence of physical effects of new entities. The classic example is the prediction of the existence of Neptune, first postulated to explain discrepancies in Uranus' orbit.  It would be hard to dismiss the proposed planet as merely an \textit{ad hoc} fix, introduced to save Newton's theory, once astronomers had it in their sights. This fits with our view that empirical evidence is the ultimate arbiter of success, even for theory-mediated inquiry.
 
In the ideal case, one can and should pursue risky predictions made by proposals for dark energy. However, the observational domain for dark energy is much more challenging to probe than either the Higgs case or Neptune. In fact, dark energy seems to combine the worst of both worlds: the far remove from unaided observation of the Higgs boson and the inability to isolate and perform experiments of Neptune. The usual arbiters of progressive versus regressive additions to a model are off the table. With dark matter, at least, there is the expectation that it is a sort of (exotic) particle. Even though it has thus far eluded direct detection, there are at least prospects for detection through non-gravitational interactions.  The case of dark energy is strikingly distinct, as we will see in the next section.  Attempts to empirically determine the nature of dark energy---whether it should be treated as a true cosmological constant, a novel particle physics effect, a result of modified gravity, or something else entirely---are by necessity quite indirect and focus on the subtle impact of dark energy in regimes such as large-scale structure formation.  

We see genuine disagreements over the best way forward regarding understanding $\Lambda$ due to these difficulties. The disagreement over the ``rules of the game'' is not so easy to resolve in cosmology as in other cases in physics, especially in the context of dark energy. In the ideal case, new physical effects are overdetermined by multiple, independent lines of evidence, so the disagreement over whether a new entity is required, whether the laws need changing, or whether the approximations made are less innocent than initially thought is transient and can be resolved. The prospects for a strong overdetermination of dark energy seem slim. We cannot say with certainty what kind of new entity dark energy is, or if it is actually an effect from new physical laws that deviate from GR at cosmic distances. In the meantime, we can assess the possible solution strategies in terms of the presupposed ``rules of the game'' for precision cosmology, pointing out ambiguity where there is room for disagreement over the best way forward.

Variation of judgements within a discipline as to the best way forward when discrepancies arise is healthy and contributes to overall progress. By exploring several incompatible alternatives simultaneously, the discipline as a whole is more likely to land on the right solution sooner. The rational strategy within the community demands variation of individual judgements. Those pursuing further elaboration of detail regarding $\Lambda$ are faced with further philosophical questions, in particular those regarding what sorts of things demand an explanation, and the types of explanation preferred or allowed. We turn to these explanation strategies in the next section.

\section{Constraining and Explaining $\Lambda$}
\label{Explanation}

Looking back on Newtonian astronomy, we see that the rules of inquiry there were well-defined: one explains minute deviations between theory and observation in terms of new gravitational sources or facts about Earth's motion, or with the aid of improved calculational techniques. With unambiguous and agreed upon rules for the game, we have clear expectations as to what types of explanation should be prioritized, and what observed phenomena require an explanation. As mentioned above, new forces or modifications to Newton's gravitational law were off limits as allowable explanations within this program, at least until all other explanations in terms of new sources or more accurate calculations were exhausted. Features not related to the motion of bodies within the solar system were not targets for explanation within this program. 

The ambiguity regarding the ``rules of the game'' for cosmology leads to differing expectations as to what sorts of things cry out for explanation and what sorts of explanations should be sought. In the case of dark energy, the relative inaccessibility of direct probes to confirm or refute a candidate explanation mean that relative priority among several explanatory virtues becomes more relevant as a deciding factor. Additionally, one's background assumptions about what modifications ought to explain accelerated expansion will be important in a physicist's preferred choice of response. In this section our goal is to map out some of the possible alternatives, and how background preferences for particular types of explanations can motivate search strategies. We begin with theoretical classes of alternatives and argue that the distinction between these types of explanation can blur in interesting ways. Due to this potential underdetermination, we argue that the best way forward is to try to make an observational case that dark energy cannot be a true cosmological constant. Even if this case could be made, we argue that there are still underdetermination obstacles as to what the correct explanation is. We make no comment on the prospects for any particular explanatory strategy; as mentioned above, it is healthy for cosmology as a whole that there are individual differences in explanatory preference, as more avenues of the solution space can be explored simultaneously. The main point of interest is that theoretical alternatives that start from different conceptual origins might be either transiently empirically equivalent, in principle empirically equivalent for beings like us, or even reformulations of the same underlying physics.

\subsection{Explanatory Strategies}
\label{ExplanatoryStrategies}

Let us begin with a brief general discussion of scientific explanation. Scientific explanation is a rich topic in philosophy of science \parencite{SEPExplanation}; our aim here is not to provide an overview of philosophical accounts of explanation. Instead, we present our own view on explanation as it pertains to understanding accelerated expansion and dark energy to illustrate the practical explanatory strategies at play. We take it that explanations are local, contextual matters; phenomena ``cry out'' for explanation when they are unexpected relative to background knowledge and fall within the scope of our theorizing, and explanations succeed insofar as they cash out in terms of other, well-understood ideas. In science, therefore, it should be clear that the process of explanation is theory-laden. Phenomena are explained in terms of the tools, concepts, and principles of theoretical frameworks, and phenomena that appear to run contrary to background expectations are the types of things that seem to demand explanation. Thus, we take it that there are no theory-independent standards for determining what phenomena should be explained and how those explanations should run.\footnote{Given the discussion in Section~\ref{Methodology}, this should not come as a surprise. Most of scientific practice is unproblematically theory-laden; attempts to give theory-neutral explanations just end up depending on more general sets of assumptions. The parameterized frameworks discussed in Section~\ref{Eliminative} are one example of this. Though they depend on more general theoretical assumptions, they do not result in some ``pure phenomenology'' completely divorced from theory.} Instead, we must situate ourselves within a scientific discipline in order to address these questions.

Generally in physics we respond to new, unexpected phenomena by positing explanations that are both risky and constrained. Anomalous results can either demand a robust physical explanation or can simply be resolved as modelling discrepancies, so it is not always the case that an empirical anomaly requires a deeper explanation. However, once a phenomenon is deemed fit for explanation, one must search for explanations that fit within the rules of the framework (constrained), but that lead to interesting new lines of inquiry that eventually produce new empirical results predicted by the framework (risky). Some of the ways that one can explain new phenomena are the following:
\begin{enumerate}
    \item[(1)] Change contingent details of the model built to account for the target phenomena, such as (a) boundary conditions or values of parameters in the model, or (b) the entities or features included in the model.  
    \item[(2)] Change the underlying dynamics used in deriving the model, such as by:  (a) incorporating new aspects of the dynamics that were initially neglected; or (b) replacing the existing dynamical laws.
    \item[(3)] Explain away the effect as an artifact by showing that it resulted from using unjustified approximation techniques, over-extending or misconstruing the model and its assumptions, or not taking systematic biases in experiment or observation properly into account.
\end{enumerate}
We start by noting that the explanatory strategies appropriate for features that appear to be necessary differ from those targeting contingent facts.  We typically aim to explain necessary features as consequences of the underlying laws, where contingent features vary among possible solutions.  This distinction can often be drawn based on repeated iterations of an experiment or manipulating a target system. In cosmology, one complication is the lack of manipulability of the target domain. Since we only have one universe (to observe, at least!) played through once, it is much harder to distinguish contingent facts about the evolution of the universe from those that are law-governed and invariant. This aside, the distinctions are at least conceptually clear, and we can use theory to help structure our expectations between necessary and contingent features. Within the context of modern cosmology, the accelerated expansion of the universe inferred from observation is a feature that can be explained by (1a) setting the contingent constant $\Lambda$ equal to the observed value, by (1b/2a) the addition of dark energy---a new type of matter field, by (2b) the modification of general relativity on cosmic scales, or by (3) explaining away the inferred accelerated expansion in some way. We can already see the sort of ambiguity regarding explanations for accelerated expansion within cosmology. Options (1a) and (1b) both involve keeping the laws and dynamics the same while changing modelling assumptions about the distribution and types of degrees of freedom. Option (1b) is more substantial, in that we include entirely new entities or features, while option (1a) involves more minor changes, such as altering the distribution of known types of matter or prioritizing other known dynamical effects. The introduction of new planets in the development of celestial mechanics exemplifies a type (1) response. The addition of dark energy is a more radical departure than adding a new planet, although (as we will see) it is contentious whether to regard it as falling under (1b), introducing a new entity, or (2a), a modification of the relevant dynamics. The exact way that one tries to frame the options can make dark energy seem innocuous or radical. As we will see by the end of this section, even the distinctions between dark energy and modified gravity might blur in nontrivial ways.

Before further exploration of types (1) and (2), we should emphasize two points.  First, there are other motivations for, and ways to assess, modified theories of gravity aside from the cosmological considerations we focus on.  A more comprehensive treatment would need to take into account other reasons for changing GR and evidential constraints from other domains.  The second point regards the importance of pursuing option (3). Attempts to explain away new anomalies play a fundamental role in the everyday process of scientific work. Theories do not come with all of their consequences explicitly worked out, and much of the work of theoretical science is in using a patchwork of theories, idealizations, and approximation methods to create a model of some phenomenon adequate to the purposes of enhancing quantitative and qualitative understanding. Given the era of increased precision in cosmological observations, extra care should be taken to ensure that idealizations, approximations, and assumptions that were fruitful or innocuous in some domains do not cause issues in new domains. We will discuss some of the ways in which classes of assumptions in cosmology can and should be tested in Section~\ref{Assumptions}. In the case where the assumptions turn out to be innocuous, we gain important insight into the domains of validity of our models and justification for continuing to trust them. In cases where they introduce unrealistic artifacts to our predictions, we can eliminate them and better understand the implications of our best theories.

As we have argued above, we think that discovering new types of matter or energy content in the universe is a legitimate aim for cosmology, exemplified by the discovery of a non-zero $\Lambda$. But it is less clear whether or not $\Lambda$ itself needs further explanation, and if so, what kind of explanation one should prefer. Is $\Lambda \neq0$ a resolution of a discrepancy in our best models of cosmology, or is it a feature that demands explanation? The consensus view seems to be that $\Lambda$ is a phenomenological paramaterization of dark energy, which demands some further explanation. This line can be explicated as follows: we have discovered a new form of energy causing the expansion of the universe to accelerate, and successful lines of inquiry should allow us to discover more about its properties. On this characterization, the demand for further understanding exemplifies the idea that new explanations must be risky. This demand for explanation can be bolstered by thinking of cosmology as parameterizing only large distance degrees of freedom; on a reductionist account of physics, we can expect this large distance behaviour to emerge from some underlying microphysics. Thus, it is also constrained by principled fits with other domains in physics. We expect to learn more about dark energy by constructing models of the underlying physics to explain what gives rise to the observed $\Lambda$ term at cosmological scales. The standard explanation in terms of a contingent constant $\Lambda$ or as $\Lambda$ in the form of quantum field vacuum energy both fall under type (1), since these involve fixing contingent constants (1a) or using known matter (1b), respectively. Treating dark energy as a new type of matter or energy falls more closely under (1b/2a), since new forms of matter are introduced. As with the program of Newtonian astronomy, postulating a new entity opens up a further line of inquiry into its properties, and ideally a direct detection. Due to the relative inaccessibility of dark energy, however, much of this work is on the theoretical side, and prospects for a direct detection are quite dim. But dark energy only works as an explanation for accelerated expansion if we have some sort of explanation as to \textit{what dark energy is}. Besides a purely contingent constant, the most conservative form of explanation is to appeal to known physics to explain the effective $\Lambda$ term. We know that vacuum energy density in quantum field theory gives rise to a term formally equivalent to a cosmological constant, so it seems like the obvious move to identify $\Lambda$ the expectation value of vacuum energy density of quantum fields.

There are two major explanatory issues that arise for treating the accelerated expansion of the universe as due to an apparent cosmological constant. First is the cosmological constant problem: a problem for quantum gravity, in which the current methods for quantizing fields leads to an anomaly---an unnatural, radiatively unstable $\Lambda$ term that is far too large to be compatible with observation. We must either explain why vacuum energy density does not gravitate or provide a mechanism to cancel out all contributions saving a small residue consistent with observations. The former option is usually called the old cosmological constant problem, while the latter is the new problem, in light of the fact that $\Lambda \neq 0$ in the $\Lambda$CDM model. Decades of unsuccessful attempts to solve this problem seem to indicate that the explanation of vacuum energy density as giving rise to $\Lambda$ is at best incomplete, at worst misguided \parencite{KoberinskiCCP,Koberinski2021,SchneiderPoS,WallaceLowEnergyQG,KoberinskiSmeenk2022}. Some new mechanism must either explain away the radiative instability of vacuum energy density or else provide an alternative explanation not involving vacuum energy.

If vacuum energy density cannot provide a good explanation for $\Lambda$, what can? Perhaps new forces, or modifications to GR, are needed to give a microphysical description of dark energy. This brings us back to options (2a) and (2b) listed above. If dark energy is something new, it could be explained via the addition of new types of matter-energy (i.e., new entities) (2a) or via modification to the dynamics of gravity (2b). As we have seen with Newtonian astronomy, adding new entities is often favoured over modifying the dynamics, as the latter is typically seen as a more drastic change to the existing line of inquiry. Quintessence models for dark energy posit a new quantum field permeating spacetime, whose properties drive the observed accelerated expansion; generalizations of quintessence models do the same \parencite[\S 2.1]{JoyceDEModifiedGrav}. However, due to the exotic nature of dark energy, many physicists also pursue option (2b), linking apparent accelerated expansion to modifications to GR. We will argue that the distinction between these two options is not as clear as one might expect.

\subsection{Eliminative Reasoning}
\label{Eliminative}

Given the explanatory impasse we currently face, it might be helpful to approach the problem of understanding $\Lambda$ from a different perspective. Rather than beginning from conceptual distinctions on the side of speculative theory, one could try to build a phenomenological framework in which to better understand the qualitative and quantitative features of $\Lambda$. Is there a strong observational case that indicates dark energy must differ from a true cosmological constant term? Maybe $\Lambda$ is simply a constant of nature for our universe. In the same way that we don't seek an explanation for the value of $G$, perhaps observations have simply allowed us to fix a nonzero value for a free parameter in the best fit spacetime solution for large-scale structure in the universe \parencite{bianchi2010all}. If we see fixing $\Lambda$ as a way to resolve a discrepancy with allowed parameters, it is not clear why $\Lambda$ itself then demands an explanation. Disentangling the issues regarding explanation laid out above would be a more pressing issue if observation forced our hand by indicating that a true cosmological constant was insufficient for understanding the various phenomena explained by $\Lambda$. Phenomenological work employing parameterized frameworks and eliminative reasoning is an attempt at building such a case. This is type (3) explanatory work seeking to quantify the degree to which observations force us to accept GR with a true cosmological constant, or whether there is significant pressure on the framework forcing us to go beyond $\Lambda$CDM in some way. Here it will be useful to briefly review the nature of the observational case for dark energy and the eliminative program.  

The most direct empirical case for a large contribution to the total mass-energy budget from dark energy comes from the magnitude-redshift relation for high-redshift supernovae.  Assuming that the FLRW models hold as a good approximation at these scales, the accelerating expansion these observations revealed implies the existence of a large effective $\Lambda$ term.  Prior to the supernovae observations, cosmologists often accepted a large non-zero $\Lambda$ due to its role in structure formation models.  But this line of argument is strikingly indirect, and assumes not only the validity of the FLRW models but also the underlying physics of structure formation, including cold dark matter.  Several cosmologists made the case that optimal parameter fits for a variety of cosmological data imply a large $\Lambda$.  These earlier lines of argument led to estimates consistent with those based on the supernovae observations.  By around the turn of the century, cosmologists routinely set the value of $\Omega_{\Lambda} \approx 0.7$, reflecting a consensus that has been further solidified since then.  It takes nothing away from the elegant work leading to this consensus to emphasize how little it reveals about the nature of dark energy itself.  Cosmologists trying to constrain the underlying physics for dark energy have often employed an eliminative approach:  after first characterizing some space of possibilities for describing dark energy, the eliminative step establishes which ones are compatible with observations.

Any inference from observed motions to the underlying stress-energy distribution requires assumptions about both the spacetime geometry and the dynamics.  All of the eliminative approaches we will discuss here use the simple homogeneous and isotropic FLRW models to represent the universe at sufficiently large scales.  This provides an essential kinematic framework for interpreting cosmological observations. One critical line of response rejects the use of FLRW models; perhaps the observations taken to support dark energy reveal only the inadequacy of these simple models.  We will postpone this question for now and return to it in Section~\ref{CCP} below. 
Note further that the use of the FLRW models is more general than the assumption that general relativity adequately characterizes the relevant dynamical features of gravity.  Many modified theories of gravity also admit the FLRW models as solutions, and in that sense the geometry of these models applies much more broadly.  Granting this kinematical background, we can then attempt to use observations to resolve two basic questions about dark energy.  First, granting that the gravitational dynamics is described by GR, are observations compatible with a true cosmological constant, or do they instead favour dark energy that varies at large scales? Second, and more ambitiously, can we determine whether observations favour introducing novel infrared features of gravity or dark energy?  To answer both questions, cosmologists have constructed a common kinematic framework based on the FLRW models to clarify the observational signatures of modified gravitational theories compared to GR plus dark energy.  

Responses to the first question treat dark energy as characterized by an equation of state that may vary over cosmic time (as a function of the redshift, $z$).  The equation of state for an ideal fluid is typically characterized by a dimensionless parameter $w = \frac{p}{\rho}$ (where $p$ is pressure, and $\rho$ energy density).  For a true cosmological constant, $w = -1$; more generally, the dark energy equation of state may vary as a function of redshift, $w(z)$.\footnote{This can also be described as the equation of state varying with cosmic time, which is well-defined in FLRW models. This generalization respects the symmetries of the FLRW models; a further generalization with $w$ varying as a function of spatial location at large scales would violate these symmetries \parencite{huterer2017}.}   Establishing that $w$ departs from a constant value would change the goal of dark energy physics; cosmologists would then need to explain not only the source of the dark energy contribution but also to account for the evolution of the equation of state. This amounts to a demonstration that some new dynamical entity is required in our model of the universe (option (2a)), since a pure constant would no longer be feasible.  Several observing programs have been designed to constrain $w(z)$ based on observations of the universe's expansion history and observables related to structure formation.  The Dark Energy Survey has constrained cosmic shear and galaxy clustering (along with many other observables) based on a sample of 100 million galaxies, both of which provide sensitive probes of $w(z)$ \parencite{Abbott_2022}; the space-based Euclid observatory slated for launch in 2023 will obtain significantly higher precision measurements of shear and other features of large scale structure.  The results obtained so far remain compatible with a true cosmological constant, frustrating efforts to identify and constrain new aspects of dark energy physics.           

Returning to the second question, since the early 2000s, theorists have explored the possibility that the observations show that we do not understand gravity rather than indicating the presence of dark energy.  No one expects GR to apply universally, since we presumably need a quantum theory to describe the gravitational interaction in some regimes (such as at high curvature). There is no analogous theoretical reason to expect the theory to fail in the infrared regime relevant for cosmology, but there is the more prosaic concern that applying GR to cosmological scales involves an enormous extrapolation from the length scales (roughly solar system scale) where it has been subjected to precision tests. It is natural to seek a theory that retains the theoretical and observational virtues of classical GR yet describes the universe's expansion history and structure formation without the need for dark energy.  Although there is more work to be done in pursuing this line of thought, the results obtained so far suggest that it is quite difficult to find such a theory.          

The alternative theories of gravity typically drop one or more of the assumptions leading to Einstein's field equations, while retaining the general idea that gravity should be represented in terms of spacetime curvature.  For example, one can treat the Ricci scalar $R$ that appears in the Einstein-Hilbert action as the first term in an expansion that includes other terms---so-called $f(R)$ theories.  Even more generally, one can include higher order terms that are a function not just of the Ricci scalar, but of other curvature invariants, $f(R, R_{ab}R^{ab}, ...)$.  This is one avenue of obtaining a more general metric theory of gravity, but there are various others \parencite[see, e.g.,][]{Clifton2012}, such as introducing further fields (as in bi-metric theories) or changing the dimensionality of space-time.  More radical modified theories take further departures from GR, including adding non-local effects or treating the field equations as an emergent property of some underlying micro-physics rather than as characterizing a fundamental interaction. Perhaps obviously, these strategies fall into type (2b) explanations for $\Lambda$. Although there are several viable alternative theories currently being investigated, it is not easy to modify GR in the infrared without generating new problems, such as instabilities associated with the new degrees of freedom.

From the observational side, it is striking how stringently results from the solar system regime constrain infrared modifications.  Most modified gravity theories have to explicitly break the link between solar system physics and the infrared regime using a screening mechanism such as the ``chameleon mechanism.'' This allows modifications of gravity to ``change colors'' in different regimes to ensure that modifications in the cosmological regime remain hidden at solar system scales.  The idea was introduced by \textcite{khoury2004chameleon}, who constructed a model with a scalar field whose mass varies directly with the local matter density.  In more dense regions (such as the solar system), the mass of the scalar field is sufficiently high to suppress interactions with matter, leading to negligible modifications of GR, whereas departures from GR become manifest at cosmological scales. There are now a number of distinct proposals to sever the connection GR provides between solar system dynamics and cosmology \parencite[see, e.g.,][]{JoyceDEModifiedGrav}.  In these theories, modifications of gravity escape notice in the solar system regime because physical features of this environment, such as relatively high density, suppress the new interactions that lead to departures from GR.  

A further difficulty is in coming up with a way to compare different models to observational results that allows for a broad space of possible models to be constrained in a simple formalism. Eliminative programs have been proposed in light of the proliferation of alternative theories of gravity.  Rather than carrying out calculations for each of the many new proposals in order to test them, this approach aims to formulate a general framework that allows one to constrain the space of possible modifications.  The hope is to elucidate a large parameter space of possible theories, such that each of the currently proposed alternative theories, as well as related ideas that have yet to be discovered, occupies a specific region of parameter space.  Furthermore, the assumptions common to all of the theories captured by the framework should be sufficient to allow the use of inferences from observations to rule out some parts of parameter space. This formalism allows for a phenomenological comparison of a potentially infinite number of possible modified gravity theories; increased precision in observations made using this formalism will either show tension with predictions from GR or will further constrain modified gravity theories in the extent to which they may deviate from GR.\footnote{This method of eliminative reasoning and repurposing precision evidence occurs elsewhere at the frontiers of physics, such as particle physics and foundations of quantum theory \parencite[cf.][]{KoberinskiFrameGen}.} We can think of this work as increasing the level of observational constraint for future inquiry; either we end up constrained to go beyond GR to understand cosmic dynamics, or we are more strongly constrained in future model building to closely match the predictions made by GR.

An influential implementation of this approach in a more straightforward case can help to illustrate the idea.  In the case of solar system dynamics, a number of competing theories can be compared to observations using a parameter space consisting of 10 parameters.  A gravitational theory maps to a set of parameters in this space, although this mapping may be many-to-one, and observations can then be used to determine which regions of parameter space are viable.  This ``Parametrized Post-Newtonian'' (PPN) approach has led to a strong observational case that GR (or a theory that is equivalent to it in this regime) gives the most empirically successful account of gravitational dynamics \parencite{Will2014}. We should also highlight two significant advantages of this approach.  We mentioned the first briefly above, namely that the space of theories captured by the parameterized framework is much broader than the list of currently proposed alternative theories.  There are a number of ``known unknowns'' that can be evaluated---that is, parameter values that may correspond to a new gravitational theory that has not yet been explicitly formulated.  Second, this approach bypasses difficult questions regarding theoretical equivalences among different proposed theories.  (This practical advantage has a downside we will return to momentarily.)  

Several cosmologists have developed a ``Parametrized Post-Friedmannian'' formalism modeled on the PPN approach.\footnote{Here we will describe the work of Tessa Baker, Pedro Ferreira, and Constantin Skordis (with several collaborators), although others have also developed eliminative programs.  In more recent work, they call their approach the ``Effective Theory of Cosmological Perturbations'' since the PPF label had already been used by \textcite{Hu_2007}.}  
This approach focuses on the dynamics governing the evolution of small density perturbations in a background FLRW spacetime.  The PPF approach parametrizes the space of possible modifications to the field equations, leading to a more general system of equations governing this evolution including 22 coefficient functions.\footnote{These coefficients are functions rather than constants, since they depend on the length and time scale of the perturbation.}  These are not all independent, and they often take a simpler form in specific regimes.  \textcite{baker2015observational} show that for large-scale perturbation modes, given some further simplifying assumptions, the effect of modifications to the underlying dynamics can be distilled into two functions. This possibility space is more complicated than in the PPN formalism, but this approach still takes advantage of the well-understood domain of linearized perturbation theory to give a comprehensive account of possible modifications.  This is still a prospective program, in at least two senses.  First, the background assumptions used in constructing the space of alternatives is not nearly as well-understood as in the PPN case.  Second, current observations only weakly constrain the parameter space, and it remains to be seen how effectively new observations of large-scale structure can do so---although obviously the formulation of the eliminative program aims to identify particularly valuable observational targets.   

This line of work promises to elucidate the constraints placed on different approaches to dark energy, granting the applicability of the FLRW models.  Yet even successfully executing an eliminative argument may not be sufficient to force a choice among various theoretical options for modeling dark energy, raising the threat of a particularly strong form of underdetermination.

\subsection{Underdetermination}



One of the main advantages of the eliminative program has an important drawback when it comes to distinguishing among different ways of understanding $\Lambda$.  Many distinct theories may map onto the same point or region in the parameterized possibility space. From the point of view of the phenomenology tracked by the eliminative program, they are entirely equivalent---any differences among the theories disappear in the regime being studied.  While the observers may be grateful that they do not need to sort out issues of theoretical equivalence to proceed, this raises an important underdetermination threat. We know both from the observational side and the theoretical side that there is some level of equivalence between different classes of models explaining $\Lambda$, and therefore there will be some level of underdetermination. Since the type and severity of underdetermination is still unknown, we survey here various forms of underdetermination discussed by philosophers. 

Transient underdetermination---where two theories are equivalent relative to a specific regime or subset of possible observations---is often not problematic, as each theory should suggest differing results in other domains that can be explored.  For example, Newtonian gravity and GR agree on descriptions of solar system dynamics up to some specified level of precision.  But this transient underdetermination can be broken either by considering strong-field gravitational effects or obtaining higher precision observations of the solar system.  Ideally, looking at complementary domains, or with greater precision, would also help to draw a sharper empirical contrast among different accounts of dark energy, in line with the account of methodology outlined above. This necessarily involves going beyond the regime of the PPF program but does not pose a problem in principle. In practice, however, it may be difficult to find empirically accessible regimes that break the underdetermination for theories of dark energy. At the most extreme, transient underdetermination may be unresolvable if possible distinguishing observations are not actually feasible for beings like us. In this case, there are possible differences between the models, but none that we will ever be able to detect due to our own limited perspective. This provides a challenge to explaining dark energy with type (1b) or (2) explanations.

Even more problematic forms of underdetermination are those that are more than transient. Persistent underdetermination between modified gravity and dark energy poses a challenge to a realist\footnote{These issues of underdetermination, theoretical equivalence, and theory interpretation are all extremely relevant for the scientific realism debate. As a rough distinction, realists are committed to the idea that our best theories faithfully (though fallibly and approximately) represent the world as it really is. The main motivating argument for realism is an inference to the best explanation: the success of our best scientific theories is best explained by their being approximately true. Antirealists are skeptical of the realist argument, and motivate their skepticism by an inductive argument over the history of science. Past theories have been successful, yet were nevertheless replaced by new, incompatible theories; by induction, we should expect our current theories to suffer the same fate. This issue has been one of the most dominant in the last 50 years of philosophy of science, and the details go well beyond the scope of this paper. For a survey, see \textcite{sep-scientific-realism}, and for an articulation of a view closest to our own on the debate, see \textcite{Stein1989}.}  understanding of either type of model. Philosophers have long discussed problems of underdetermination as a challenge to literal interpretations of theories \parencite{Duhem,Quine1951,Quine1975,Stanford2006}. Some have argued that, in principle, there are always an infinite number of logically possible ``theories'' compatible with all of the evidence at any given time, and that one could create similar theories that make all of the same predictions as the one currently accepted by the scientific community. The conclusion that these philosophers have drawn is that literal interpretation of theories as accurate representations of the way the world \textit{really is} is untenable, since any base of evidence can equally support many logically possible theories. Responses to these in principle issues have been dismissive: despite the logical possibility of multiple theories predicting all the same evidence, in practice this is not something that typically happens in science. Building even one empirically equivalent adequate theory, let alone two equivalent theories, is usually enough of a challenge. In the philosophical literature, critics have argued that there are extra-empirical reasons to prefer some theories over others that are equally compatible with a body of data. Simplicity, beauty, understanding, unification, and other criteria have all been posited as epistemic values that license us to give higher credence to one theory over another. Some of these are epistemic features of fruitful research programs, while others are aesthetic considerations which therefore might be thought of as more dubious means of justification. A second important response strategy is to argue that supposed underdetermination should indicate that the theories are actually equivalent in important ways.  One version of this response reconsiders what we should count as empirical content, and then makes the case that theories sharing content, so defined, should be regarded as equivalent \parencite{Wilson1980}. Alternatively, many concrete examples of rival theories can be reinterpreted as being mere notational variants of the same theory, where one can establish a straightforward sense of equivalence \parencite{deHaro,Weatherall2019}.


If we move beyond observational equivalence for transient underdetermination, but we stop short of admitting any logically equivalent competitors, then dualities between models and reformulations of the same theory are more realistic forms of underdetermination leading to theoretical equivalence. Dualities are conjectured translations between seemingly quite distinct theories, such as the AdS/CFT duality, whereas reformulations are different mathematical representations of the same physical domains, such as canonical quantization and the path integral formalism. Depending on the type of underdetermination between dark energy and modified gravity and the way one chooses to interpret each formulation, theoretical equivalence could be cast in terms of duality or reformulation. Even if a robust form of theoretical equivalence can be established, pressure is placed on any explanation that involves a straightforward reading of any mathematical formalism as representational \parencite{Bokulich2020,Fraser2020}. Philosophers intending to interpret theories in physics have often sought to avoid privileging one formulation of a theory over others; if the formulations are different mathematical representations of \textit{the same} underlying physical reality, then that reality should be interpretable in terms of all possible reformulations. Simple examples of this interpretive rule include treating gauge degrees of freedom as unphysical or reformulating Newtonian spacetime into Galilean spacetime to provide the minimal spacetime structure needed for the theory.

While this rule seems reasonable, one can see how it puts pressure on the distinction between dark energy and modified gravity in cases where there is a full theoretical equivalence between models. If accelerated expansion can be equivalently described by models with a prima facie distinct physical interpretation, what grounds do we have to privilege one over the other? If we have no grounds for privileging one, how are we to interpret the underlying physics? In cases where the underdetermination is transient, this is a less pressing concern. We can resolve the underdetermination by pursuing further inquiry to differentiate among theories that agree within a specific regime. Even if the transient underdetermination is in practice irresolvable, the models would not be fully theoretically equivalent, and there may be good reasons to prefer one model over the other, such as coherence with other theories, fruitfulness for other avenues of inquiry, or unification. The most stringent challenge comes when the underdetermination is in the form of a full equivalence or duality between models.

Some modelers have tried to take advantage of possible ambiguities by constructing models that have aspects of both dark energy \textit{and} modified gravity, allowing the interpretation to change in different regimes. \textcite{MartensLehmkuhl} discuss the blurred distinction between spacetime and matter in the context of a dark matter model, concluding that the distinction between the two may not be as stark as has been supposed. Even in the simple case where dark energy is an effective cosmological constant, partially sourced by vacuum energy density of quantum fields, the distinction between matter and spacetime is blurred. The expectation value of vacuum energy density is supposed to be constant throughout space (at a given cosmic time),\footnote{This is standardly assumed when setting up the cosmological constant problem, but it is far from obvious why local applications of Minkowski quantum field theory should be extrapolated to cosmic distances \parencite{KoberinskiSmeenk2022}.} and its effects therefore could be thought of as part of the background spacetime structure, on top of which quantized excitations propagate. However, it is only the presence of matter that provides this permeating effect; take away the matter, and the cosmological constant would also change. But we also know from GR that spacetime is dynamical, and that removing matter from the universe changes the spacetime geometry. So the distinction is already quite blurry with the most obvious explanation for accelerated expansion, let alone from more sophisticated models.

\textcite{JoyceDEModifiedGrav} provide a thorough survey of dark energy versus modified gravity models and propose a means of observationally distinguishing between them. They categorize the distinction heuristically as follows: dark energy models alter the stress-energy content of the universe---the right-hand side of the Einstein equations---while modified gravity models alter the gravitation and spacetime content of the equations---the left-hand side. Since it is mathematically permissible to move terms to either side of the equality, this can only be a rough heuristic: further physical differences must be present to distinguish between the two. The diagnostic Joyce, et al. use to distinguish between dark energy and modified gravity is the strong equivalence principle: they classify any models that continue to satisfy the strong equivalence principle as dark energy models, and those that do not are treated as modified gravity models. If we take the universality of free-fall to be a necessary defining characteristic of GR, then it makes sense to think of models that violate this principle as somehow modifying GR, rather than introducing new matter-energy sources. An example is the class of scalar-tensor theories where the Einstein-Hilbert action is modified by the addition of a new scalar field $\phi$, including its kinetic energy and some potential $V(\phi)$. Rather than coupling to the metric, matter couples to a function $A(\phi)g_{\mu\nu}$, with $g_{\mu\nu}$ the usual GR metric. If $A(\phi) \neq 1$, then the effect of the new scalar is to modify the gravitational force and to decouple it from the geometry, violating the strong equivalence principle. If $A(\phi) = 1$, then the new scalar field does not alter the gravitational force between other types of matter, and interacts in some other way; the strong equivalence principle is upheld, and we can think of $\phi$ as a new energy source for dark energy. 

Again, this is a clear way of formulating the distinction between adding new entities versus changing the laws, similar to the means we propose. Joyce, et al. reach a similar conclusion to the one we draw here:
\begin{quote}
while there are models which unambiguously belong to one category or the other, in reality there is a continuum of models between the two extremes of “pure” Dark Energy and Modified Gravity such that a strict division into these two categories is to some extent a matter of personal preference (p. 2).
\end{quote}

One of the things that makes dark energy interesting philosophically is its resistance to fitting into the standard categories laid out by philosophers studying other areas of science. Even within physics, the standard ``rules of the game'' only provide a launching point; explanations for remote, esoteric phenomena like that parameterized by $\Lambda$ seem to blur the conceptually clear distinctions that were useful in guiding inquiry at the frontiers of nineteenth century physics. What counts as an addition of new entities versus a modification to dynamics is not clear cut; nor do those distinctions align perfectly with the ideas of modifying contingent versus necessary facts.

\section{Typicality assumptions}
\label{Assumptions}

Typicality assumptions play a central role in the practice of contemporary cosmology. Even the minimalist eliminative programs described above assume the applicability of FLRW geometry, and this is usually justified by appealing to the uniformity we observe in conjunction with an assertion that our vantage point is ``typical'' in an appropriate sense. While it is impossible to pursue any line of inquiry without background assumptions, one can check the self-consistency of models by critically examining the tenability of the principles that went into constructing it (type (3) explanation from Section~\ref{Explanation}). In this section, we discuss the various forms of typicality assumptions and assess their use in cosmology. These assumptions and principles have many features in common, though their status differs. The Copernican principle and cosmological principle aim to characterize precisely different senses in which we are typical and clarify what follows from this. \textcite{Carter1974}'s seminal discussion of anthropic principles starts by noting that we can err in \emph{overstating} the Copernican idea that we do not occupy a privileged position.  Clearly creatures like us can only exist in a very specific physical environment.   
The line of thought we will pursue here is mostly focused on formulating typicality assumptions and their implications, while acknowledging this crucial point. 
As we will argue below, we do not regard the ``strong'' anthropic principle or related discussions of fine-tuning invoking a multiverse as similarly well-founded.\footnote{There is a vast literature on anthropic principles and fine-tuning, from the classic early overview \textcite{barrowtipleranthropic} to the more recent collection \textcite{carr2007universe}; from more philosophical discussions we recommend \textcite{Earm87,Roush03,friederich2021multiverse}. Our treatment here is not a survey; we instead argue for what we take to be justified and unjustified uses of typicality assumptions in cosmology.}   
 The forms of typicality reasoning that work are well-supported by empirical consistency checks and provide a basis for fruitful inquiry, while those that fail are not sufficiently grounded in empirically plausible assumptions, running the risk of circularity. Finally, we reconstruct the use of anthropic arguments in the ``prediction'' of $\Lambda$ and the role they played in motivating natural solutions to the cosmological constant problem. We start with some general comments on the necessity of typicality assumptions in science.

Typicality assumptions are not limited to cosmology, though they tend to be more controversial here due to the relative inaccessibility of the target system---for some purposes, the entire observable universe. Scientific inquiry nearly always requires typicality assumptions of some form, as the use of statistics in testing hypotheses illustrates. We must assume that the results of repeated measurements are typical of the phenomena that we aim to investigate. We can always test assumptions about selection bias in our data, but this amounts to arguing that the measurements we make are representative of a different class of phenomena than we originally thought. This sort of typicality assumption is a necessary starting point for inquiry, justified through its use. Further, the assumption could turn out to be false.\footnote{This is a more practical version of \citeposs{Lewis1980} Principal Principle, though stated in terms of probabilities rather than credence: the probability of outcome $A$ should be equal to the (objective) chance of $A$:
\begin{equation}
Pr(A | Ch(A) = x) = x.
\end{equation}
Even assuming this principle, for any fundamentally chancy events, we could just turn out to be extremely unlucky; the probability of getting 100 heads in a row on a fair coin is small, but never zero.} Our data might fail to be typical; perhaps the act of setting up an experiment heavily skews the outcomes in ways that are not representative of the actual chances for the target events. However, we cannot learn from evidence without assuming that the evidence is representative in some sense. The better-motivated typicality assumptions in cosmology are of this kind. In looking for selection bias, scientists standardly assess whether we have chosen the appropriate class of phenomena for which our evidence should be treated as representative. Some checks on the plausibility of the cosmological and Copernican principles are type (3) explanations of this sort. It can turn out that we need to revise this class, but the assumption that the evidence is not representative of \textit{some} more general class of phenomena is a retreat to a radical form of inductive skepticism. 

Even those core methodological principles that are well-justified and stand up to repeated scrutiny may turn out to be false. We argue that one should think of these typicality assumptions in their strongest form as  regulative principles (in the spirit of \textcite[I.114]{Peirce}). These are guiding principles that are needed in order to get a productive line of inquiry off the ground and are foundational to a discipline. Regulative principles are often shielded from refutation, though their foundational necessity is not a form of evidence for their truth; they could turn out to be false. They may be required to underwrite a certain line of inquiry, but perhaps that is a line of inquiry we simply cannot pursue. 

We will argue below that the typicality assumptions in cosmology should be treated as regulative principles in this sense.  In cosmology these issues are often discussed in terms of a bewildering variety of principles:  Copernican, cosmological, strong and weak anthropic, and so on.  One of our tasks below will be to clarify the content of these principles and their interrelations.  We will take the Copernican principle, often formulated as the claim that ``we are not privileged observers,'' as the starting point, but making this vague claim precise leads to a number of distinct ideas.  The cosmological principle, for example, makes a precise claim regarding the absence of privileged locations with regard to spacetime geometry. 
The weak anthropic principle is useful as a way to highlight selection bias, revealing that our evidence is not representative of the reference class we initially thought, but instead some other class of phenomena biased by our epistemic location. We argue that the weak and strong anthropic principles actually serve entirely different purposes, such that the naming convention is misleading. One \textit{could} characterize the strong anthropic principle as stemming from the Copernican principle, but these two principles differ dramatically in whether their use generates further evidence. Further, the strong anthropic principle makes use of vague reference classes of ``agents in the multiverse'', within which our observations are meant to by typical. Since the theoretical background assumptions behind strong anthropics are neither well-supported empirically, nor susceptible to consistency checks of type (3) (Section~\ref{Explanation}), we argue that they are unjustified. Closely tied to strong anthropic arguments are issues with defining probability distributions over reference classes of observers. Finally, we discuss the anthropic ``prediction'' of $\Lambda$ and argue that it was a good theoretical motivation to search for a solution to the cosmological constant problem, but does not provide evidence for any range of values of $\Lambda$.

\subsection{The Copernican and cosmological principles}
\label{CCP}

Typicality assumptions can be justified by empirical consistency checks and their role in supporting further inquiry. They vary in their level of testability, but most are at least indirectly testable. The regularity of the fundamental laws across time and space is on one extreme of a typicality assumption that is difficult to test but necessary for empirical science. On another extreme are the fine-grained specific modeling assumptions of an experimental design, such as the sufficiency of the sample size or the photometric calibration. In this class are the many statistical rules we regard as generally trustworthy and whose applicability in a given experiment can be determined through independent tests.

The Copernican Principle posits that our position as observers in the Universe is typical. Tests of this assumption can involve checks on the appropriate class of phenomena regarded as typical, i.e., in order to ask where in the Universe are observers \textit{like us} likely to be, we need to specify what observers \textit{like us} are. The Copernican Principle is sometimes phrased as ``we don't live in a special place in the Universe,'' or that ``there are no privileged observers.'' Obviously we cannot change our location in the universe to see how things look from elsewhere and elsewhen to directly test its truth.  In addition to this epistemic limitation, these familiar formulations remain vague until we make ``special locations'' or ``privilege'' precise.

Postponing the question of its content for a moment, it is clear that the Copernican Principle has an important status as a background assumption in contemporary cosmology.  Some version of the principle has been assumed by most cosmologists throughout the last century, and even though it might be false it would not be easily discarded---we expect cosmologists would give up a number of other assumptions in order to save it. But rather than treating it is an a priori principle, we suggest that its status is that of a Peircian regulative principle.  But there is no guarantee that the principle holds: the world could be such that an appealing line of inquiry is simply not feasible.  The local void models considered in cosmology, discussed below, nicely illustrate this point: grant for the sake of argument that it is possible to construct cosmological models of this type, compatible with the data, for which the typicality assumptions fail---we would be observers in a special location, close to the center of an extremely large void.  On our view, the failure of typicality assumptions does not immediately rule these models out.  But cosmologists who lived in such a universe would only be able to pursue radically different goals. They would not be able to use precision cosmological observations to discover general facts about what kind of matter and energy dominate large-scale dynamics---their observations would instead merely reveal features of the local environment.  

A regulative principle is a weaker notion than the perhaps better-known Kantian proposal, transcendental principles \parencite{Kant}. Transcendental arguments start from some accepted aspect of experience, taken as given, and then show what must be the case in order for those aspects of experience to be possible. A transcendental argument for the Copernican principle would seek to establish its truth from its necessity for cosmology:  first, argue that precision cosmology is possible, then establish that some form of Copernican principle is necessary for the possibility of precision cosmology. Using the concept of a regulative principle instead, we can grant both aspects of this argument without granting the conclusion; it could very well be the case that the Copernican principle is false, but some form of it is necessary to for us to use observations from our vantage point to draw conclusions about the large-scale structure of the universe as a whole. Nature does not have to conform to our desire to do science.

Without a transcendental argument for its truth, how can we justify the Copernican principle as a typicality assumption? Most philosophers of science, from Carnap to Kuhn, would argue that there can be no purely objective, rational justification for this sort of regulative principle. Its ultimate justification is in its successful application in a productive scientific discipline. As \textcite{smith2014closing} argues, successful science must be predicated on a theoretical framework, including a number of principles and assumptions. The ability to discover new facts in the world using the background framework ultimately justifies the use of that framework. Even if the framework turns out to be incorrect or in need of revision, the stable facts in the world it allows us to discover give us epistemic justification for its use (within some restricted domain). Often, the only way to discover that the framework needs revision is by assuming its truth and discovering persistent discrepancies between observation and prediction. Our suggestion is that the best justification for the use of the Copernican principle in cosmology takes precisely this form.

As we noted above, there is further work to be done to clarify the content of the principle, to make claims about ``special locations'' or ``privileged positions'' precise. The Copernican principle is often discussed in the same breath as the cosmological principle, which admits a sharp mathematical formulation:  it states that the Universe is \textit{homogeneous} and \textit{isotropic} at sufficiently large scales. The term goes back to Milne, who regarded the cosmological principle as an a priori axiom for his cosmological theory.  It is no longer popular to treat this as an a priori axiom, yet it still plays a foundational role in defining the basic spacetime geometry used in the $\Lambda$CDM model:  in the context of general relativity, stipulating that homogeneity and isotropy hold leads directly to the extremely simple geometry of the FLRW models. One can treat the cosmological principle as a stronger and more precise form of the Copernican principle: in a truly homogeneous and isotropic universe, there are no special places or privileged observers \textit{anywhere}. We have removed some of the vagueness of the Copernican principle by specifying homogeneity and isotropy as the conditions of typicality. 

Alternatively, we can take homogeneity and isotropy to follow from the Copernican principle in conjunction with observations from our own vantage point---such as the isotropy of the CMB.  A line of results starting with \textcite{EGS} take the following form:  if fundamental observers moving along a congruence of timelike geodesics through a region $R$ observe collisionless radiation that is exactly isotropic, granting some further plausible assumptions about the matter distribution, the spacetime geometry of region $R$ is given by an FLRW model.  The Copernican principle is needed to make the step from observations of isotropy along one curve to a claim about a congruence of fundamental observers, but then observed isotropy can be leveraged to establish the cosmological principle.  Various results along these lines have been established with progressively weaker assumptions regarding the matter distribution and the observed isotropy (see \cite{EMM} \S 13.1).     

Our initial formulation of the cosmological principle has the virtue of precision but the vice of being false. We know from observation that it cannot apply on planetary, solar system, or even galactic scales, where we observe inhomogeneity and anisotropy. Clearly it needs to be qualified in some way, to be treated as a statistical claim regarding ``small enough'' departures from uniformity at ``large enough'' scales:  e.g., the mean density within any sphere of radius $R$ approaches the mean density of the Universe for large enough $R$. This is more challenging than it might at first appear due to the difficulty with defining ``averages'' in a generic spacetime:  there isn't a background spacetime we can use to define volumes.  More generally, we need to address what is sometimes called the fitting problem:  namely, how well do cosmological observations determine a ``best fit'' FLRW model \parencite{ellis1987}.  We also need to understand the limits of using these simple models, as emphasized in the discussion of type (3) tests in Section~\ref{ExplanatoryStrategies}.  Easier to formulate than Copernican claims of non-specialness, tests of homogeneity and isotropy allow the definition of a precise scale at which these approximations apply: observations of the cosmic microwave background (CMB) define a scale of isotropy (at the time of last-scattering), and observations of large-scale structure can potentially define a scale of homogeneity \parencite{Hogg2005,Goncalves2018}.

There are two broad types of questions we can ask when examining the modeling assumptions and approximations of our theoretical framework---in this case, the cosmological principle \parencite[see also][]{Smeenk2020}. Firstly, in what ways might the cosmological principle fail? We can probe the validity of the assumptions of homogeneity and isotropy in several ways:
\begin{itemize}
    \item Do they hold in the early universe?
    \item Do they hold today? 
    \item At what length scales do they hold?
    \item Are they only partially valid, e.g., is the Universe \textit{only homogeneous} or \textit{only isotropic}?
\end{itemize} 
Secondly, what are the effects of de-idealizing the cosmological principle on our observational reports? This is crucially important for explanations relating to the cosmological constant, given that we know the cosmological principle fails below some scale (let's say 50-100\hmpc), and again, this question can take several forms:
\begin{itemize}
    \item What is the effect on our observations of light propagation in an inhomogeneous universe?
    \item What is the effect of peculiar motions with respect to the cosmological ``Hubble flow'' due to the dynamical effect of local inhomogeneity?
    \item What is the effect of averaging over local inhomogeneities to arrive at our observational parameters---the cosmological back-reaction?
    \item What is the magnitude of these effects, and can any, or a combination, of these effects of local inhomogeneity explain the accelerated expansion of the Universe?
    \end{itemize}
Different lines of inquiry will focus on one or a few of these (and other) questions at a time, examining the de-idealization of the cosmological principle along a particular line of inference. 
It is particularly difficult, however, to determine the effect of dropping the assumption of homogeneity and isotropy because that involves developing analytical models of nonlinear perturbation theory within GR. Numerical simulations of large-scale structure formation are heavily relied on for this reason \parencite[see, e.g.,][]{Falck2021}, and they work in the Newtonian limit of GR with homogeneous and isotropic evolution equations \parencite{Peebles1980}. 
We briefly consider two classes of attempts to develop exact analytic or numerical cosmological solutions to the field equations which allow some degree of anisotropy or inhomogenity: models of a more realistic lumpy universe that do not violate the Copernican principle and the so-called ``local void'' models that do.

The first class of models we consider are those that do not violate the cosmological or Copernican principles; rather, they seek to work out the details of how these principles apply, and at what scales, given observations of small-scale inhomogeneity and anisotropy. These work firmly within the theoretical framework of GR and attempt to de-idealize the exactly smooth FLRW metric by adding perturbations. In some models, this can lead to extra ``back-reaction'' terms in the evolution equations that influence the expansion history of the universe \parencite{Buchert2011,Clarkson2011}.
In the simplest ``Swiss cheese'' models, perturbations are modeled as spherically-symmetric underdensities distributed throughout a smooth background, with the advantage of being exact solutions to Einstein's field equations and the disadvantage of being a less-realistic approximation of the actual distribution of matter \parencite{Kantowski1969,Fleury2014,Koksbang2017}. 
Whether or not the magnitude of cosmological back-reaction terms or the distance-redshift relations calculated in Swiss cheese models can explain the apparent accelerated expansion, these lines of inquiry are an example of the important work of checking the consistency and working out the implications of the modeling assumptions within the theoretical framework. 

In the second class of models, the nonlinear perturbations are not necessarily uniformly distributed, resulting in an inhomogeneous universe which may violate either or both of the cosmological and Copernican principles. For example, some models place us at or near the center of a very large spherically symmetric void, as opposed to somewhere within a uniform distribution of voids. In some models, a local underdensity is required with a diameter greater than 400\hmpc\ or even 2\hgpc, well outside estimated upper limits on the homogeneity scale (see \textcite{Goncalves2018} and references therein). Models with such large central voids are generally understood to violate the Copernican principle, whether because of their size or because of the necessity for us to be at or near the center of the void. These types of models can be easily constructed to fit the distance-redshift relation \parencite{Garcia-Bellido2008,Celerier2000}, but they have a more challenging time reconciling with observations of structure formation and the CMB. To the extent that they can be tested empirically, they explore type (3) explanations by checking the cosmological typicality assumption of the Copernican principle.

The Copernican and cosmological principles concern the typicality of observations that relate to our location in the Universe, the first stating that we are not in a special location, the second that there are no special locations on very large scales. But we can't sample different locations in the Universe and repeat our observations to determine whether they are typical. Indeed, a random location in the Universe has a much higher probability of being in a low-density void than in or near galaxies: by volume, any massive object is in a ``special'' place in the Universe. On the other hand, by mass, the centers of massive galaxies are the ``typical'' places. Whether a typicality assumption holds depends strongly on the phenomenon being considered; asking whether a measurement is representative of a phenomenon is a way of comparing what is typical \textit{for the measurement} to what is typical \textit{for the phenomenon}. In the case of the Copernican principle, the phenomenon is ourselves, human beings. Where in the Universe are entities like us likely to be, and is our place typical of those environments? Placing these kinds of constraints falls under anthropic reasoning.

\subsection{Weak anthropic principle}
\label{WAP}

Many different principles get grouped together under the umbrella of ``anthropic principles''. 
We will use the standard contrast between ``weak'' and ``strong'' versions of the anthropic principle going back to Carter, although on our view this nomenclature blurs the contrast between two quite different types of reasoning. The differences are important: while weaker, more modest anthropic arguments can play an important heuristic role in all areas of science, stronger forms are unjustified and should be rejected.  
We start by examining the modest arguments, which we formulate below, without any claim to originality, as the ``weak anthropic principle.'' 
These arguments will help us get a handle on the sorts of things anthropic reasoning in general can and can't accomplish and provide a contrast with the common features of strong anthropic arguments. At core, we think of anthropic principles as making explicit the often implicit premise that a constraint on our best theories of the universe is that they must be compatible with our existence and observational abilities at our specific location in space and time. (In some sense, we can think of this as schematizing the observer within our theories \parencite{stein1994some,Smeenk2020}.) 
Justified uses of anthropic principles are all of a relatively weak form: they can be used to explain sources of selection bias in observation or measurement, or to point out flaws in the inference chain from observational constraints to justified theories. With these legitimate, but deflationary, uses of anthropic principles in mind, we will criticize the use of strong anthropic principles in Section~\ref{SAP}.

In the simplest terms, the weak anthropic principle is the (perhaps tautological) claim that the universe is compatible with our existence at a certain location within it. This puts constraints on our theories and on the inference chain from observation to phenomena. For the former, we can think of the weak anthropic principle as providing consistency constraints on our theories, which may motivate specific lines of inquiry, while for the latter, we can think of it as bringing attention to possible selection effects in our experiments. Selection effects explain away something that might seem puzzling or unlikely from the point of view of our current theory and known background assumptions. Highlighting a selection effect is making explicit a previously unconsidered background fact that might alter the force with which something cries out for explanation.

There are many selection biases in astronomy. For example, the well-known Malmquist bias describes how our observations of object distances are affected by the limits on our observations of brightness due to the correlation between apparent brightness and distance. When we infer the distribution function of object brightnesses---the luminosity function---from the measurements of a magnitude-limited astronomical survey, the result is both an effect of the phenomenon of interest -- how many objects of a given brightness there are in a given volume---and of the process that produced the measurements---the noise properties of the telescope, the calibration steps, and many other effects, including selection effects. The Malmquist bias is limited in its explanatory scope: it explains (features of) the \textit{observed} distribution rather than the actual distribution of the underlying phenomenon; it explains why and in what way the observed and actual distributions differ. We must take it into account in order to determine the actual distribution, but the Malmquist bias makes no claims to explaining why astronomical objects are as bright as they are. Selection effects are necessary links in the chain of inference from observation.

Selection effects are usually thought of as a sampling bias: the selected sample does not represent a random sample of the underlying population. We can find sampling bias in our observations or experiments, but we may also find that selection effects block inductive inferences from local measurements to global features of gravity or spacetime. For example, (an unconditional form of) the cosmological principle might imply that local measurements of mass-energy density are good indicators of overall mass-energy density on cosmic scales. But one can easily point out the selection effect here: we live in a region of very high mass-energy density compared to large-scale averages, so a local measurement is not representative of the global quantity. These are anthropic in an indirect sense: our local environment as observers, or our situatedness in a specific spacetime location, are enough to explain the selection effect. 

At its most general, the weak anthropic principle can appeal simply to the fact that we exist. This is not a form of typicality assumption: instead, it is a form of conditional reasoning by modus ponens. \emph{If} the universe is hospitable for beings like us to exist, and our best supported physical theories are correct, \emph{then} this set of parameters $\{A_i\}$ must be bounded in the interval $\Delta$; The universe is hospitable to beings like us; \emph{Therefore} $\{A_i\} \in \Delta$. Modus ponens is a well-established inference rule of deductive logic, and invoking it in this form is justified. If we are justified in taking the premises to hold, then we are justified in believing the conclusion. Conversely, if we get evidence that the conclusion is false, we must revise the premises. Since we take our existence to be fixed, this would imply that some aspect of our best theories is false, or that we have erred in the derivation of $\{A_i\} \in \Delta$ from our best theories. We can think of this as an application of the principle of total evidence: in order to fully understand the implications of our observations and how they relate to a theory's predictions, we need to keep in mind \textit{all} relevant and available evidence \parencite{White2000}. Our existence and situatedness in a specific location in the universe is sometimes relevant evidence. But note that on this way of thinking, there is nothing specifically \emph{anthropic} that is essential:  we could reach the same conclusions by noting the existence of, well, literally anything that exists. Replacing ``beings like us'' in the first premise with anything else (a flower, a rock, or a neutron star) might change our derived bounds on the parameters but would not change the logic of the argument.

The weak anthropic principle, when used legitimately, is meant to bring attention to a selection effect: the allowed values of $\Lambda$ or other fundamental constants are constrained by the nature of how they are observed. Historically Carter (and before him, Dicke) were both inspired to consider anthropic selection effects due to various apparent coincidences in the values of large numbers constructed from fundamental physical constants.
If we expect typicality with respect to some theoretical background without taking selection effects into consideration, we can be misled into thinking that such coincidences demand further explanation. In cosmology, weak anthropic arguments can remind us to reconsider the appropriate class of phenomena of which our observations of the universe can be expected to be typical. However, the weak anthropic principle does not \textit{explain} the nature of the laws or principles that lead to constants taking the allowed range of values. Before discussing anthropic arguments applied to dark energy and $\Lambda$, we discuss unjustified uses of anthropic arguments next.

\subsection{Strong anthropic principle and the multiverse}
\label{SAP}

The conventional nomenclature is misleading:  the strong anthropic principle is a different type of principle rather than a strengthened version of what we have just described.  It plays two distinct roles: first, providing a connection between a multiverse theory and what we observe; and, second, as an explanation for some set of observations in terms of a selection effect across pocket universes in the multiverse. We argue that both of these uses of strong anthropic reasoning are problematic, and further, that positing a selection effect in the multiverse cuts off the search for a further explanation for $\Lambda$, in much the same way as treating it as a contingent constant. That does not mean the multiverse account can't be true; rather, as a methodological move, it also gives up on seeking a deeper physical account of $\Lambda$. We note that many different formulations of the strong anthropic principle exist in the literature; the differences are not relevant for the discussion here. Instead we focus on critiquing the common argumentative structure that they share; insofar as any anthropic argument shares in this structure, it is subject to the critique below.

Strong anthropic reasoning typically takes the following form, with three types of assumptions needed to generate claims regarding the expected value or range of permissible values of some fundamental physical parameters (including, but not limited to, $\Lambda$). On a traditional hypothetico-deductive account of scientific explanation, a successful argument of this kind provides a candidate explanation for the values of those parameters. First, we assume a physical theory that allows the values of fundamental constants to vary, either counterfactually (in distinct possible but not actual universes) or actually (in distinct regions of a single causally connected universe, usually called the multiverse). Several well-established theories allow counterfactual variation of the constants, but currently theories that apparently generate \textit{actual} variation in distinct regions of the multiverse remain speculative. We will call the assumption that one of these theories, such as string theory or eternal inflation, holds ``speculative physics'' (\textit{SpecPhys}). We will state the existence of the multiverse itself as an explicit second premise, \textit{Multiverse}, given that there are ongoing debates regarding whether specific fundamental theories in fact generate a multiverse. What turns this argument into a form of anthropic reasoning is the third premise, \textit{Anthropic},  that ``observers like us exist'' (or, in some formulations, \textit{must} exist). From these three assumptions we can derive the conclusion that the parameter values $\{A_i\}$ in our universe must fall within a certain range $\Delta$ compatible with our existence. Schematically, this is just a conditional claim:\footnote{Here $\wedge$ denotes logical conjunction, $\vee$ logical disjunction, and $\rightarrow$ the conditional.}
\begin{equation}
\label{eqSAP}
    (SpecPhys \wedge Multiverse \wedge Anthropic) \rightarrow \{A_i\} \in \Delta.
\end{equation}
What should we make of a stronger claim: that observation of $\{A_i\} \in \Delta$ provides evidence for the combination of $(SpecPhys \wedge Multiverse \wedge Anthropic)$? Logically speaking, this is just the fallacy of affirming the consequent. There could be many possible mechanisms that lead to $\{A_i\} \in \Delta$, so until we have explored the space of competitors and ruled them out there is no positive evidence for a multiverse or the speculative physics models. However, observing $\{A_i\} \notin \Delta$ would count as disconfirming evidence for $(SpecPhys \wedge Multiverse)$, since we can hardly avoid granting that we are here to ask these questions. At best the ``prediction'' of parameter values provides a sort of consistency check on the speculative physics \parencite{Smeenk2014}. So the prospects for using strong anthropic reasoning to provide evidence for speculative models in physics, in the form of predictions for the values of fundamental constants, is tenuous.

It is more plausible to take \eqref{eqSAP} as providing an explanation of the observed values if we have independent reason for accepting the antecedent conditions. This is where the problem lies, however. Starting with the premise $Multiverse$, what independent grounds could we have for it? For other types of self-location selection effect arguments, we can observe that there are other possibilities, e.g., other locations in spacetime, other solar systems, other galaxies, etc. But we have no telescopes or other instruments that give us access to other pockets of the multiverse.\footnote{There have been some attempts to obtain indirect evidence for multiverse theories, such as traces left over in the CMB of collisions between different bubbles in the early universe. Even if traces with the appropriate signature were to be found, which they have not, this would still be quite weak evidence, unless other plausible physical mechanisms for generating such effects could be ruled out.} In fact, most multiverse advocates treat $Multiverse$ as a consequence of $SpecPhys$ rather than an independent postulate, and the strength of the explanation then depends entirely on $SpecPhys$. As the name we have chosen suggests, we do not currently have the kind of empirical justification for candidate fundamental theories that might entail a multiverse that we have for other physical theories. At best, then, we could consider the strong anthropic explanation as a possible candidate explanation, contingent on the success of a given model of speculative physics that entails the existence of a multiverse.

But it is not clear to us what role the multiverse itself would play in such an explanation.  Suppose that physicists arrive at a well-supported theory, according to which a nondeterministic, stochastic process in the early universe fixes parameters like $\Lambda$, as in a model of eternal inflation or a symmetry breaking process from string theory. In what sense does a mechanism like this entail a multiverse? Or, to put it another way, what work is the multiverse posit doing in the explanation of the value of the parameter? A stochastic mechanism might only run once; the move to assuming that a probability distribution over possibilities implies an ensemble of actual outcomes does not seem to add anything to the explanation. To give a concrete example, consider electroweak spontaneous symmetry breaking in the context of the evolution of the universe. As temperature drops, the electroweak symmetry breaks, and there is a continuous infinity of possible, inequivalent new vacuum states. Unless one interprets quantum theory along Everettian lines, this is a one-off event perfectly well-described as a stochastic process; our universe could have been in any of the other vacuum states, though it settled in this one. It appears as though we have a clear understanding of the different possibilities without appealing to a multiverse where every possible vacuum state was occupied, and whose probability is proportional to the measure over the space of universes. We do not see any reasons that we \textit{must} interpret this situation in terms of a multiverse, with distinct realizations of the phase transition in different regions.  Why should other stochastic mechanisms be any different?  

On our view, the claim that the multiverse plays an essential explanatory role here reflects a different conception of anthropic reasoning:  namely, that our presence as observers acts as a selection effect over an actually existing ensemble of universes. 
In this way, it would be closer in form to the weak anthropic principle. However, in this form one must treat the multiverse as already well-established if one aims to treat such an argument as explanatory or predictive for the value of constants like $\Lambda$. We will return to this distinctive understanding of anthropic considerations in the next section, but the difference in style of reasoning in the two cases strikes us as much greater than a contrast between ``weak'' and ``strong''.  

Much more deserves to be said about the strong anthropic principle and the line of reasoning described above, but we will limit ourselves to two further remarks.  First, many versions of anthropic reasoning attempt to assign probability distributions over the values of fundamental parameters appearing in Eq.~\eqref{eqSAP}.  Doing so would have obvious benefits, as it would support moving from a mere consistency check to assigning probabilities or degrees of credence to competing fundamental theories.  But introducing probabilities in this context faces conceptual and technical challenges \parencite{freivogel2011making,Smeenk2014}, 
such as the well-known ``measure problem'' in eternal inflation. Second, an example due to \textcite{Aguirre2001} shows that we should be cautious about the whole conditional Eq.~\eqref{eqSAP}.  Aguirre constructs a universe with a completely different set of cosmological parameters that apparently satisfies the requirements usually imposed for a universe to be ``hospitable to life''---the existence of large gravitationally-bound systems like galaxies, complex chemistry, long-lived stars, etc.  This suggests that there may be several distinct, non-overlapping regions $\Delta, \Delta', \Delta,'' ...$ in a higher-dimensional parameter space that are hospitable to life, rather than a single region centered on the actual observed values.  If the conditional in fact implies that $\{A_i\} \in \Delta \vee \Delta' \vee \Delta'' \vee ...$, as Aguirre's example suggests, this further weakens the claim that the argument could provide evidence in favor of the multiverse.

Returning to our main line of argument, there is an important sense in which the proposed strong anthropic ``explanation'' for $\Lambda$ as an environmental effect cuts off fruitful lines of inquiry. If the only explanation is that $\Lambda$ is contingent and has no further dynamical connection to deeper physics, we have cut off a line of inquiry that seeks to understand $\Lambda$ in terms of dynamics, laws, or other necessary features of the universe. In a sense, this is similar to the \textcite{bianchi2010all} argument, except the strong anthropic principle relies on unconfirmed, speculative physics to make the point. Like the purely contingent constant case, this explanation could turn out to be true, but it seems to us premature to stop inquiry into $\Lambda$ on the basis of speculative physics.

\subsection{Anthropic bounds on $\Lambda$ and naturalness}
\label{FTN}

In thinking about anthropic arguments for $\Lambda$, Weinberg's famous ``prediction'' placing bounds on $\Lambda$ immediately comes to mind. Using the general discussion of the weak anthropic principle, we can reconstruct the reasoning in its historical context here. We argue that in its most justified form, the weak anthropic principle was used to highlight a conflict between the assumption that both the standard model of particle physics and general relativity were correct and compatible, and the more secure assumption that the universe must be hospitable to life. This conflict could be taken as motivation for finding a natural solution to the cosmological constant problem, which would resolve the conflict by correcting our theories.

At the time of \citeposs{Weinberg1987} prediction, there was little direct evidence constraining possible values of $\Lambda$, given what else we knew about cosmology, so there was not a strong conflict between observation and the large value of $\Lambda$ ``predicted'' by the conjunction of the standard model and GR. Instead, Weinberg supplied an argument for the conflict between a large value of $\Lambda$ and a seemingly unrelated empirical fact---the existence of life in the universe. By holding fixed our knowledge of structure formation and assuming that gravitationally bound systems are necessary for the formation of life, we get an upper bound on permissible values of $\Lambda$, $\Lambda_{max}$. This provides a tension with the assumption that the standard model is correct, and that vacuum energy density sources $\Lambda$: for any reasonable cutoff scale, $\Lambda_{SM} \gg \Lambda_{max}$. This is the cosmological constant problem. Given the wealth of evidence for the standard model and GR, this prompts one to search for a solution to the problem that retains the successes of our theories but changes the value of $\Lambda_{SM}$. This was part of the dominant program in beyond standard model particle physics for decades: searching for \textit{natural} solutions to the cosmological constant problem \parencite{Weinberg1989}. The role that the anthropic principle played here was in generating conflict between theory and experience when we did not have direct evidence for any value of $\Lambda$. Recent observational constraints on $\Lambda$ provide evidence of the conflict that is more direct, since this evidence is less mediated by inferential assumptions about values of parameters and their impact on structure formation.

Naturalness is another type of typicality assumption. At a basic level, naturalness is the expectation that dimensionless numbers in a theory should be of ``medium-size''---that is, not too much larger or smaller than 1. We can think about this as an expectation of what sorts of numbers are typical in nature. As stated, this is both a poorly justified principle and one that is not born out in the practice of physics. The fine-structure constant, for example, is $\alpha \approx e^2/4\pi\epsilon_0\hbar c \approx 1/137$, a number smaller than one by more than two orders of magnitude.
        
Behind the idea of unnatural, fine-tuned parameters is often an expectation that certain parameter values are more likely than others. This is an unjustified typicality argument in cases where we don't have a handle on the dynamics (if any) governing the parameter values. Without this handle, ascriptions of probability are unmotivated and unjustified. Probability claims are something we achieve after study, not the sort of thing that can be assumed \emph{a priori}. Even if something like a preference for medium-sized numbers made sense, some have argued that naturalness issues in particle physics---in the form of the Higgs mass problem and the cosmological constant problem---are an artifact of poor regularization conditions on the effective theories, akin to choosing bad coordinate systems against which to renormalize \parencite{Manohar2020,Rosaler2019}. It is at least not clear how justified naturalness assumptions of this form are and whether they should be used as guiding principles in further inquiry, especially given the seeming intractability of the cosmological constant problem as an explanation for dark energy.
        
However, the idea of naturalness has been developed to a more justified extent. Rather than fine-tuning, one can cash out the idea of naturalness in terms of symmetry groups (technical naturalness), or even the assumption that physics at distant scales decouples. These more precise notions are better justified but typically of more limited scope. \textcite{Williams2015}, for instance, traces the common origin of naturalness assumptions in high-energy physics to an assumption of decoupling: that physics at one energy scale is largely independent of physics at far distant energy scales. This is a notion of rather limited scope, since the precise definitions of decoupling here apply to local, Lagrangian field theories, whose high-energy formulation must be renormalizable \parencite{Appelquist1975}, though the guiding principle of decoupling is meant to generalize this to all effective field theories. \textcite{Wallace2019} makes a strong case that this form of naturalness is foundational to the way we currently do physics and to how we understand theoretical relations like emergence within physics. From Wallace's perspective, at least, it would appear that naturalness plays a role similar to the Copernican principle as a sort of regulative principle for physics. If this is the case, then a natural solution to the cosmological constant problem gets further motivation from the assumption that this regulative principle holds. \textcite{Giudice2008} can be read as saying something similar regarding naturalness, as a ubiquitous assumption as to what sorts of scaling behaviour are typical. However, there are still two major outstanding failures of naturalness for the standard model of particle physics: the cosmological constant problem and the Higgs mass problem. Whether thought of as fine-tuned parameters, parameters not protected by a higher symmetry group, radiatively unstable terms, or failures of decoupling, these two terms stick out as exceptions to any expectations of naturalness. \textcite{Giudice2017} has since argued that empirical results from the LHC indicate that naturalness is no longer a good guiding principle for fruitful lines of inquiry into physics beyond the Standard Model. \textcite{KoberinskiSmeenk2022} also argue that one should expect decoupling and the effective field theory program in which it is formulated to break down in the context of cosmology. Naturalness therefore seems to be an obsolete typicality assumption even in the context of particle physics. Therefore, it may no longer be useful for understanding $\Lambda$.
        
Even if a precise form of naturalness \textit{was} a good guiding methodological principle in the past, its failure to bear empirically fruitful models should be considered as a failed critical test of this foundational principle. \textcite{Giudice2017}, a former champion of naturalness as a guiding principle, has argued that particle physics has entered a post-naturalness era. The role that anthropic arguments played in motivating the solution strategy was historically important but has since been superseded by direct empirical conflict. Further, the failure to find an empirically successful natural extension of the standard model should at least motivate the idea that naturalness might break down here. This is an interesting type of unjustified assumption, in that it is largely due to (lack of) empirical results that the assumption fails, rather than some flaw in reasoning.

\section{Conclusions}
We have argued for a perspective on modern cosmology as a largely successful, fundamentally iterative enterprise. Using guidance from general relativity, cosmologists have discovered many details of the evolution of our universe, from fractions of a second after the big bang up to today. A cosmological constant $\Lambda$ parameterizes an essential ingredient of the resulting $\Lambda$CDM model, without which our understanding of structure formation and accelerated expansion would fail. Though we currently understand little about the nature of $\Lambda$, and there are several interesting open problems as to the best way forward, we think that minority claims of crisis in cosmology are misguided.\footnote{To be clear, the claims about a crisis we have in mind are those based on objections to including a non-zero $\Lambda$; there may of course be other reasons, such as the $H_0$ tension, for concluding that all is not well with $\Lambda$CDM.} When we view cosmology as a fundamentally temporally extended line of inquiry, constrained by known details but risky in generating independent lines of empirical evidence, we see the reintroduction of $\Lambda$ into cosmology as a legitimate move. However, the legitimacy of $\Lambda$ in the $\Lambda$CDM model does not resolve a host of issues about the methodology of cosmology, the goals of explanation, and the possibility of underdetermination of exactly what dark energy is. Part of what makes $\Lambda$ so challenging is its remoteness from empirical control: it is both unobservable and impossible to isolate in the lab.

Whether $\Lambda$ demands an explanation, and if so what type of explanation should be preferred, are open questions that depend on one's (perhaps unstated) philosophical background assumptions. We know little about $\Lambda$, and the prospects for direct detection are slim, so many theoretical alternatives offer competing explanations that fit the evidence equally well. As of now, all evidence is compatible with treating $\Lambda$ as a true cosmological constant. One prospect for progress is the observational program of constraining the equation of state of dark energy, or using the parameterized post-Friedmann formalism to search for deviations in structure formation from that predicted by $\Lambda$CDM. These eliminative programs aim to provide direct observational evidence for dark energy or modified gravity by establishing that it cannot be fully captured as a cosmological constant term. If this program is unsuccessful, we get stronger and stronger evidence to treat $\Lambda$ as a true constant; if successful, we have discovered new features of the universe, to be explained by new physics. This represents progress in finding new empirical avenues for conducting precision tests of dark energy. One major worry is that these explanations will be strongly underdetermined. Despite these underdetermination worries, such a project furthers the lines of inquiry into the dynamically relevant details of the universe, and there is no reason to think that the underdetermination cannot be resolved by continued inquiry. Even if we hit a dead end, we will hopefully have made progress in understanding more about the universe by pursuing these types of explanation.

Like any scientific enterprise, modern cosmology is predicated on a set of theoretical principles and assumptions. Some of these are deemed necessary for the project of cosmology, some are testable within cosmology, while others might be found to be unjustified and untestable. The necessity of some principles---like the Copernican principle---is not evidence for their truth, but the success of lines of inquiry predicated on these regulative principles supports their continued use. The bounds of applicability of the cosmological principle can be tested by modelling alternative solutions in which it fails on various scales; these tests have implications for $\Lambda$ insofar as the inference from observation to $\Lambda$ relies on isotropy and homogeneity. Models where the cosmological principle fails either fail to match predictions in other regimes from $\Lambda$CDM, or else to not have enough of an effect to explain away $\Lambda$ as an artifact of overidealization. The value in this work is that it provides stronger evidence for the validity of modelling practices in cosmology.

Explanations of $\Lambda$ that invoke typicality assumptions across a multiverse are of a more dubious character. Unlike the Copernican principle, they do not seem necessary for the project of modern cosmology, and unlike the cosmological principle they are not independently testable in any way. It is also not clear how explanatory they are: in the end, the value of $\Lambda$ for our universe is a contingent fact, in much the same way as the bare posit that $\Lambda$ is just an observationally determined constant. Given the extra speculative baggage accompanying strong anthropic explanations, it seems preferable to adopt the latter explanation. The most justified use of any anthropic reasoning regarding $\Lambda$ was in the historical context of the cosmological constant problem, as a way to illustrate that a large value of $\Lambda$ seemingly predicted from particle physics is grossly incompatible with the presence of structure in the universe, given what else we know of physics. This motivated the search for natural mechanisms to solve the cosmological constant problem, but these seem to us separable from the observational case for $\Lambda \neq 0$. The failure of the naturalness program for particle physics should also motivate a search for a different type of explanation.

Dark energy in modern cosmology brings together many interesting issues in the philosophy of science in ways that are often not found in other contexts. We think that fruitful discussion between philosophers of science and cosmologists can enrich both fields and offer new perspectives on progress in cosmology. By thinking of cosmology as a risky and constrained line of inquiry, a framework for thinking about ways forward for $\Lambda$ emerges, as well as an understanding that the community should explore multiple lines of explanation. We also see that competing theoretical explanations will likely face underdetermination issues of a kind more pressing than usually considered in philosophy of science. Overall, modern cosmology is in a very good position, with prospects for ever more precise observation to increase our understanding of the details that make a difference in the evolution of our universe. Discrepancies and transient tensions can often be the strongest source of evidence over the long term.

\section*{Author Contributions}
Conceptualization, Adam Koberinski, Bridget Falck and Chris Smeenk; Investigation, Adam Koberinski, Bridget Falck and Chris Smeenk; Methodology, Adam Koberinski, Bridget Falck and Chris Smeenk; Writing---original draft, Adam Koberinski, Bridget Falck and Chris Smeenk; Writing---review \& editing, Adam Koberinski, Bridget Falck and Chris Smeenk.

\section*{Acknowledgments}
We would like to thank three anonymous reviewers and Sofie Koksbang for helpful comments on a previous draft.
\printbibliography

\end{document}